\documentclass[debug]{rmaa}

\usepackage{paralist}

\usepackage{psfrag,color}

\usepackage[latin1]{inputenc}

\title{The Effect of Sublimation Temperature Dependencies on Disk Walls Around
T Tauri Stars}

\author{
  Erick Nagel,\altaffilmark{1} 
  Paola D'Alessio,\altaffilmark{2}
  Nuria Calvet,\altaffilmark{3}
  Catherine Espaillat,\altaffilmark{4}
  and Miguel Angel Trinidad\altaffilmark{1}}

\altaffiltext{1}{Departamento de Astronom\'\i a, Universidad de 
Guanajuato, M\'exico.}

\altaffiltext{2}{Centro de Radioastronom\'\i{}a y Astrof\'\i{}sica,
  UNAM, M\'exico.}

\altaffiltext{3}{Department of Astronomy, University of Michigan, USA.}

\altaffiltext{4}{Harvard-Smithsonian Center for Astrophysics, USA.}

\shortauthor{Nagel et al.}
\shorttitle{Sublimation walls around T Tauri stars}

\fulladdresses{
\item Nuria Calvet: Department of Astronomy, University of Michigan, Ann 
  Arbor,MI, 48109, USA (ncalvet@umich.edu).
\item Paola D'Alessio: Centro de Radioastronom\'\i{}a y
  Astrof\'\i{}sica, Universidad Nacional Aut\'onoma de M\'exico,
  Apartado Postal 3--72, 58090 Morelia, Michoac\'an, M\'exico
  (p.dalessio@crya.unam.mx).
\item Catherine Espaillat: Harvard-Smithsonian Center for Astrophysics, 
  Cambridge, MA, 02138, USA (cespaillat@cfa.harvard.edu).
\item Erick Nagel and Miguel Angel Trinidad: Departamento de Astronom\'\i a, 
  Universidad de Guanajuato, 36240, Guanajuato, Guanajuato, M\'exico 
(erick@astro.ugto.mx,trinidad@astro.ugto.mx).}

\listofauthors{E. Nagel, P. Dalessio, N. Calvet, C. Espaillat, \& M.A. Trinidad}
\indexauthor{Nagel, E.}
\indexauthor{D'Alessio, P.}
\indexauthor{Calvet, N.}
\indexauthor{Espaillat, C.}
\indexauthor{Trinidad, M.A.}

\abstract{The dust cannot survive closer to the star from the point where a 
grain reaches a temperature equal to the sublimation temperature. 
The boundary between a dust-free and a dusty region defines the
sublimation wall. 
In the literature two models for the structure of the wall are used: a
wall with a fixed sublimation temperature and a wall with a density-dependent
sublimation temperature. In the former, the wall is vertical and in the latter,
the wall is curved.
We find important differences between these models SEDs in the wavelength range
from $3$ to $8\mu$m, being the emission 
of the former larger than that of the latter model. We quantify
the differences in IRAC colors when these models are used to explain the 
observations. In the IRAC color-color diagram ([3.6]-[4.5] vs. [5.8]-[8.0]),
the models are located in specific regions, either depending on the 
inclination, the mass accretion rate, or which model is used. 
}
 
\resumen{El polvo no puede sobrevivir m\'as cerca de la estrella de un punto 
donde alcanza una temperatura igual a la temperatura de sublimaci\'on.
La frontera entre una regi\'on sin-polvo y polvosa define la pared de 
sublimaci\'on. En la literatura dos modelos para la estructura de la pared 
son usados: una pared 
con una temperatura de sublimaci\'on fija y con una temperatura de
sublimaci\'on dependiente de densidad. En la primera, la pared es vertical y
en la segunda, la pared es curva. Encontramos diferencias importantes
entre la SEDs de estos modelos en el intervalo de longitudes de onda desde
$3$ a $8\mu$m, siendo la emisi\'on de la primera m\'as grande que la segunda.
Cuantificamos las diferencias en los colores de IRAC
cuando estos modelos son usados para explicar las observaciones. En 
el diagrama de IRAC color-color
([3.6]-[4.5] vs. [5.8]-[8.0]), los modelos est\'an localizados en regiones
espec\'\i ficas, dada la inclinaci\'on, la tasa de acreci\'on
de masa, o cu\'al de los modelos es usado.}

\addkeyword{Infrared: general}
\addkeyword{Protoplanetary disks}
\addkeyword{Stars: Pre-main sequence}

\begin{document}
\maketitle

\section{INTRODUCTION}
\label{sec-intro}

One of the motivations for the study of the dust in the protoplanetary disks 
inner regions is the fact that terrestrial planets are formed there 
\citep{1973ApJ...183...1051,2010Ast...10...19}. The dust might not survive 
if the dust grains reaches temperatures higher than the sublimation temperature.
Thus, the innermost region of a disk around a star is dust free and has an
opacity hole or deficit. The outer boundary of the hole formed in this way is 
called a sublimation wall or the disk inner rim. This surface separates an 
outer dusty and an inner gaseous disks, and its shape depends on the 
characteristics of the gas and the dust. 

\citet{1992ApJ...397...613} model the near-infrared (NIR) excess of Herbig 
Ae/Be stars as coming from material located at the inner rim of the disk with a
temperature around $1500K$. This is consistent with the evaporation temperature
of silicate grains, thus, they interpret that the emission is produced in a 
sublimation wall. In fact, the edge of the hole produced by dust sublimation 
mostly emits in the NIR. The modeling of 
the NIR excesses as produced in this kind of walls was followed by many 
\citep{2001Nat...409...1012,2005ApJ...623...952,2005ApJ...635...1173,
2005ApJ...624...832,2007PPV...539}. A nice characterization of the shape of
the wall using a Monte Carlo radiative transfer code, taking into account
stratification of grains was developed by \citet{2004ApJ...617...1177}.
They found an empirical formula 
for the radius of the sublimation wall ($R_{wall}$) in terms
of the evaporation temperature $T_{sub}$ ($R_{wall}\propto T_{sub}^{-2.085}$), 
which is consistent with simplified expressions commonly used. 
The range of dust sublimation temperature in terms of gas density is between
$10^{-18}$ and $10^{-4}g\,cm^{-3}$ and for different grain species is given by
\citet{1994ApJ...421...615}. For the silicate olivine, the range is 
$(929-1774)K$; for the silicate pyroxene, $T_{sub}=(920-1621)K$; for troilite, 
$T_{sub}=680K$; and for iron the range is $(835-1908)K$.

For the case of very low luminosity systems as the brown dwarfs,
\citet{2010MNRAS...409...1307} notes that the location of the dust sublimation 
wall is
equal to the disk co-rotation radius with the magnetosphere, as it is also 
suggested by \citet{2007ApJ...669...1072}. For this case and also for 
configurations
where the sublimation wall is inside the co-rotation radius, the magnetic 
field truncates the disk, thus, the resulting shape of the wall depends 
strongly on this processes and weakly on the sublimation phenomenon. In 
our case, the stellar luminosity is large enough to move the wall outside the 
magnetosphere, thus, its geometry is given by the physical state of dust and 
gas. 

Taking a unique sublimation temperature, leads to a unique radius, thus, a
vertical wall. We refer to this model as $T_{0,fix}$ wall.
For this model, we use a sublimation temperature
characteristic of the disk midplane, and assume it is the same at every 
height (in spite of the variation of density with height). This implies that
the wall is vertical, i.e., the inner surface of a hollow cylinder , but also, 
that the surface temperature is constant (and equal to the assumed sublimation 
temperature). 
On the other hand, taking into account the dependence of the sublimation 
temperature with density, the wall is curved. We refer to this model as 
$T_{0,\rho}$ wall.
The variation of density with height and radius in disks models are taken from
structures assumed to
be in vertical hydrostatic equilibrium. Again, there is a twofold effect,
affecting the solid angle of the visible portions of the wall and also,
the surface temperature of the wall, which is different at each pixel.
When we compare the SED of a $T_{0,fix}$ wall with that of a $T_{0,\rho}$ wall, 
both effects are present, i.e., differences in area and in temperature, but 
them are difficult to disentangle.

If the gas density and the dust composition do not depend on the vertical 
coordinate, then a $T_{0,fix}$ wall is formed. We want to point out the fact 
that sometimes due to the unknowns of the composition and density profile, one 
is naturally leaded to simplify the system and assume
a homogeneous vertical distribution of matter. 
However, as we will see in the following, a vertical stratified model,  
curves the wall. Previously, the assumption was that the shape of the wall was 
vertical \citep{2001ApJ...560...957,2005ApJ...621...461}. 
However, for the modeling of the stationary state of the disk around EX Lup
\citep{2009A&A...507...881}, a 
$T_{0,fix}$ wall was unable to explain the IR observations, thus an {\it ad hoc} 
rounded inner rim was required. A non-vertical wall was also considered by 
\citet{2005A&A...438...899}, they 
interpreted a $2\mu$m emission bump, in observed SEDs of Herbig Ae stars, as
coming from a sublimation wall. They consider that due to the
density dependence on the sublimation temperature, the wall is curved. Two 
years later \citet{2007ApJ...661...374} pursues this further, taking into 
account grain sedimentation.
The dust is composed of two grain size distributions, characterized by different
scale heights. In these works a $T_{0,\rho}$ wall model can reproduce the Herbig
Ae 
stars NIR spectrum. \citet{2005A&A...438...899} and \citet{2007ApJ...661...374}
study the differences between synthetic images of $T_{0,fix}$ and $T_{0,\rho}$ 
wall models, here, we analyse the effects on the colors. Unfortunately the 
spatial resolution and, limited 
sensitivity and the number of telescopes of NIR interferometric observations, 
make difficult to confirm the geometry of the inner region 
\citep{2010ARA&A...48...205}. Thus, we 
have some intrinsic degeneracy on the models used to interpret the data. Due to
this, it is important to include all the physics that we can on the models of 
these walls. 

In this work, we consider that sublimation is the mechanism responsible to 
produce the inner hole, the physics involved in this process shapes the 
wall. Note that a recently formed planet \citep{2004ApJ...612...L137} and a 
photodissociation 
flux \citep{2001MNRAS...328...485} are also able to create a hole. 
The magnetic rotational instability is responsible to create a wind, which in
turn is able to form a hole \citep{2010ApJ...718...1289}. A binary system is
another way to create a hole, in this case by gravitational interactions 
\citep{2005ApJ...628...L147,2007ApJ...664...L111,2010ApJ...708...38}. The
physical peculiarities of each process will define the structure for the wall. 

In a sense, this work follows the steps of \citet{2005A&A...438...899}, because
we are taking into account the same physics
to describe the grain sublimation. However, one difference is that we focus on 
T Tauri instead of Herbig Ae stars. Our first aim is
to compare models of $T_{0,\rho}$ and $T_{0,fix}$ walls, characterizing 
parameters like
location and surface temperature. Another questions to address are: if we
change the inclination or the mass accretion rate, still is possible to 
discriminate between a $T_{0,fix}$
and a $T_{0,\rho}$ wall based only on the SED?, based on the Infrared Array 
Camera (IRAC, on board the Spitzer Space Telescope)
observations, does a model of a disk plus a $T_{0,fix}$ wall or a disk plus a
$T_{0,\rho}$ wall show differences in the IRAC colors? 
Noteworthy, \citet{2005A&A...438...899}
mention that based on an image is easy to discriminate between them, because
for the $T_{0,fix}$ one, only the half part of the wall that is farthest away
from the observer show emission in the line of sight. The observed $T_{0,\rho}$ 
wall 
emission comes from every azimuthal angle of the wall. Without an image, one
cannot choose between both models, however, looking the differences in the SED
is a way to argue if for the problem at hand it is enough to take a $T_{0,fix}$
model. 

The minimum location of the wall (where the maximum temperature occurs)
is a parameter that changes when using $T_{0,fix}$ or $T_{0,\rho}$ wall models, 
one can conclude in the following sections that this parameter differ at most 
by $10\%$ (see Table~\ref{table-rad-temp}). In the case of the IRAC colors, 
between the $T_{0,\rho}$ and $T_{0,fix}$ wall models, the [3.6]-[4.5] and 
[5.8]-[8.0] colors changes in around $20\%$ and $10\%$, respectively.
In \S~\ref{sec-code} we present the details of the code used in this work, 
followed in \S~\ref{sec-walls} by the resulting characteristics of the walls,
either the $T_{0,fix}$ (\S~\ref{sec-vertical})or the $T_{0,\rho}$ walls 
(\S~\ref{sec-curved}). \S~\ref{sec-color} shows the IRAC colors associated to
the models presented. Finally, in \S~\ref{sec-summary} we present a summary
and the conclusions.

\section{DESCRIPTION OF THE CODE}
\label{sec-code}

The $T_{0,fix}$ wall emission is calculated with the code used in 
\citet{2005ApJ...621...461} for an isolated star or for a binary system in 
\citet{2010ApJ...708...38}. We consider that the
wall is optically thick but the emission from an optically thin atmosphere is 
taken into account. The wall is heated by the impinging geometrically diluted
stellar radiation flux coming from the photosphere of the star and from the 
accretion shocks.
We assume that the stellar radiation is plane parallel. Thus, for a 
$T_{0,fix}$ wall, the radiation arriving at each point is the same.
The total emission is the addition of the 
contribution of each layer at given $\tau$, extinguished with the material in 
front of it. The temperature is calculated following \citet{1991ApJ...380...617}
and \citet{2005ApJ...621...461}.

We have assumed that the opacities are independent of $\tau$, in accordance of
\citet{1991ApJ...380...617},\citet{1992RMxAA...24...27} and 
\citet{2005ApJ...621...461}. This is a necessary 
assumption in order to find an analytical expression for $T(\tau)$. The 
temperature at every depth of the wall atmosphere is lower than the sublimation
temperature. Thus, there is no sublimation of dust, and it is safe to take a 
constant opacity.

The geometrical effects producing shadowing of regions of the wall by regions 
of the wall closer to the observer are taken into account. The SED 
of the wall is calculated integrating the flux of 
every point in the wall whose normal has a component in the direction of the 
observer, times the solid angle subtended by every pixel. 

The minimum ($a_{min}$) and maximum size ($a_{max}$) of the grains are 
$0.005\mu$m and $0.25\mu$m, respectively; the power law exponent is $-3.5$;
which are parameters typical for interstellar grains 
\citep{1977ApJ...217...425}. The dust is composed of 
silicates (pyroxenes, $Mg_{0.8}\,Fe_{0.2}\,SiO_3$; and olivines, $Mg\,Fe\,SiO_4$
), graphite and troilite. We adopt a dust-to-gas mass ratio for the silicates,
$\zeta_{sil}=0.0034$ \citep{1984ApJ...285...89}; for the graphite, 
$\zeta_{grap}=0.0025$; and for the troilite, $\zeta_{troi}=7.68\times 10^{-4}$.
The composition and abundance are typical for accretion disks 
\citep{1994ApJ...421...615}. The optical properties of the silicates come from
\citet{1995A&A...300...503}. From the dust composition chosen, the 
silicates are the grains with the highest sublimation temperature. From this
follows the fact that the silicates rules the location and shape of the wall.
Dust species with a higher sublimation temperature as corundum ($Al_{2}O_{3}$),
in principle will affect the location and structure of the wall, because such
grains are the ones formed closest to the star \citep{2006A&A...447...311}. A
consistent modeling of a disk with corundum requires a study that at the same
time takes into account the gas and dust opacity, because 
in the temperature range where corundum is formed, the gas contribution to the
opacity is a sizable fraction of the total opacity. In the region where the
silicates grains are formed, the silicates opacity is around 6 orders of 
magnitude larger than the gas opacity \citep{2005ApJ...623...585}, thus, it is 
not necessary to include the gas contribution to the case treated here. This is
the reason why in the sublimation wall formation is not important to know the 
gas opacity, something that we cannot leave aside when corundum is included in 
the mixture.

A near-IR emission study of a disk-star system should include the 
contribution of a gaseous disk inside the sublimation wall, the importance of 
this is highlighted in the interpretation of interferometric observations by
\citet{2008ApJ...677...L51} and \citet{2010ApJ...718...774}. Either for the 
study of the dust-free gaseous disk emission or the shaping of the wall by the
presence of corundum, a detailed knowledge of the gas opacity is necessary.
This is a non trivial issue, because of the presence of millions of lines and
also, because it is not clear what kind of mean opacity is representative for
the approach used here. Due to this,
the gas emission problem is beyond the scope of this paper, but should be
taken into account in the future.

The $T_{0,\rho}$ wall emission is calculated based on the code just
described, but including an arbitrary shape for the wall. Unlike a $T_{0,fix}$ 
wall, for the $T_{0,\rho}$ wall, the radiation arriving to different places of 
the wall is not the same. The impinging flux depends on the angle of incidence 
$\alpha$, (the angle between the normal to the wall surface and the incidence 
ray), specifically is proportional to $\cos\alpha$, which
in turn depends on the wall shape. Thus, we have to characterize
$\alpha$ in order to get a value for the temperature in the wall. In other 
words, we need to know beforehand $\alpha$ to get the temperature, but we 
require the temperature to calculate $\alpha$, this means the wall shape.
A way to solve this problem is to note that if the scattering of the stellar
radiation is 
neglected along with the heating from inner regions (viscously produced), an 
expression for $T(\tau=0)$ without dependence on $\alpha$ 
is found. The contribution of the scattered emission has the characteristic
frequency range of the stellar radiation, thus, its main contribution is at 
wavelengths smaller than the peak of the emission of sublimation walls. Because
of this, to neglect the scattering is reasonable when one focus in
the infrared frequency range. 
Using this expression, the wall shape is defined as the points where this
temperature is equal to the sublimation temperature. This turns out in a shape 
for the wall surface; from where an incidence angle can be calculated. 
With these assumptions, the temperature as a function of $\tau$ can be written 
as:

\begin{equation}
  T(\tau)^{4}={(L_{\star}+L_{acc})\cos\alpha\over 16\sigma\pi r^{2}}
(c_{1}+c_{2}e^{-{q\tau\over \cos\alpha}}),
\label{eq_t}
\end{equation} 

where $q=\kappa_{inc}/\kappa_{d}$, 

\begin{equation}
  c_{1}={3\cos\alpha\over q},
\label{eq_c1}
\end{equation}

and

\begin{equation}
  c_{2}={q\over \cos\alpha}-{3\cos\alpha\over q}.
\label{eq_c2}
\end{equation}

Then, the temperature at the surface ($\tau=0$) is

\begin{equation}
  T(\tau)^{4}={(L_{\star}+L_{acc})\cos\alpha\over 16\sigma\pi r^{2}}
\left({3\cos\alpha\over q}+\left({q\over \cos\alpha}-{3\cos\alpha\over q}\right)
e^{-{q\tau\over \cos\alpha}}\right),
\label{eq_t1}
\end{equation} 

and we know that it should be equal to the sublimation temperature,
$T_{sub}(\rho(z,R))$ \citep{1994ApJ...421...615}, which depends on density. 
This results in an equation for $\alpha(z,R)$.

Here, $\kappa_{inc}$ and $\kappa_{d}$ are the mean opacity of true absorption, 
evaluated at the temperature characteristic of the incident stellar radiation,
and at the temperature of the disk, respectively. $L_{\star}$ is the luminosity
of the star and $L_{acc}$ is the luminosity produced by the shocks of the 
material accreting along the magnetic field lines. Note that $q$ depends on the
temperature, thus equation~\ref{eq_t1} is solved 
iteratively.
Knowing the temperature, the radius of dust destruction is calculated 
substituting $\tau=0$ in equation~\ref{eq_t1}, thus,
\vskip\baselineskip

\begin{equation}
  r_{des}(z)^{2}={L_{\star}+L_{acc}\over 16\pi\sigma}\left({\kappa_{inc}\over
\kappa_d}\right){1\over T_{sub}(z)^4},
\end{equation}
\vskip\baselineskip

in which the scattering and the local radiation field are neglected.
\vskip\baselineskip

\section{CHARACTERISTICS OF WALLS}
\label{sec-walls}

The differences in the SED when comparing a $T_{0,\rho}$ and a $T_{0,fix}$ wall, 
are the result of a combination of, at least, 2 effects: 1) geometry, because
in a $T_{0,\rho}$ wall, each pixel shows a different effective area to the 
observer than in a $T_{0,fix}$ wall  and 2) surface temperature gradient, because
what we are assuming that curves the wall is the dependence of the
sublimation temperature with density, thus this implies that at
each height, the surface of the wall would have a different temperature
than a $T_{0,fix}$ wall, defined with a unique sublimation temperature.

\begin{figure}[!t]
  \includegraphics[width=\columnwidth]{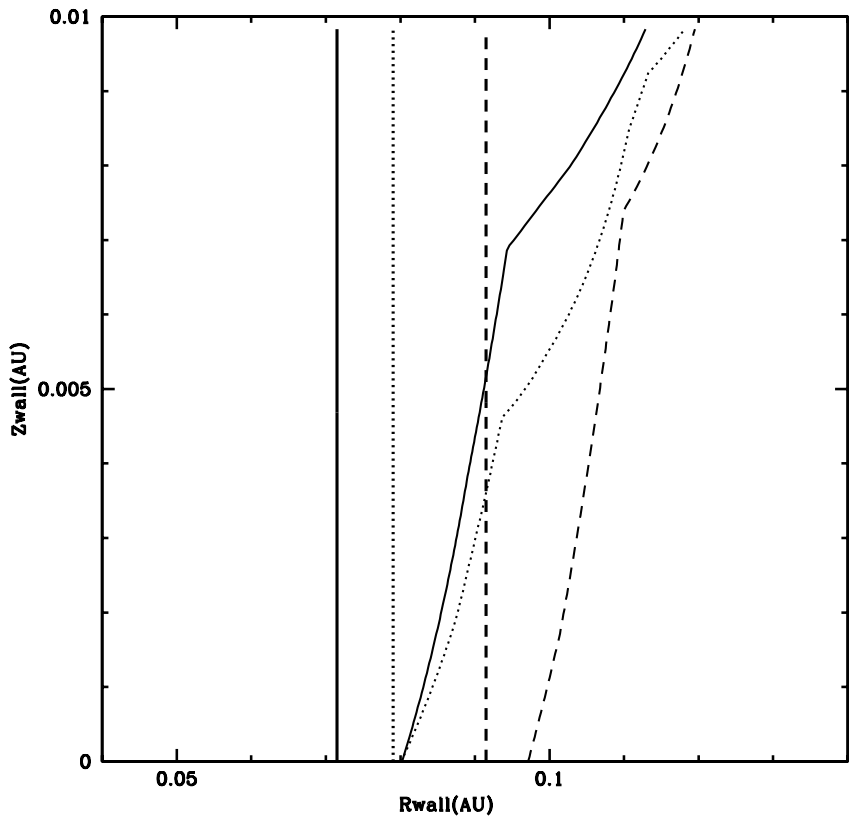}
  \caption{The shape of the $T_{0,fix}$ walls are shown with thin lines.
The $T_{0,\rho}$ walls shapes are presented with thick lines. Models presented 
are $\dot{M}=1.625\times 10^{-8}M_{\odot}yr^{-1}$ (solid line),
$\dot{M}=3.25\times 10^{-8}M_{\odot}yr^{-1}$ (pointed line), and  
$\dot{M}=6.5\times 10^{-8}M_{\odot}yr^{-1}$ (dashed line).
}
\label{fig-shape}
\end{figure}

In this section, we calculate the wall emission using the code described in 
\S~\ref{sec-code}. Our intention is to construct models for the wall emission 
for a typical young low mass star: $M_{\star}=0.5\,M_{\odot}$, 
$R_{\star}=2.0\,R_{\odot}$,$T_{\star}=4000K$, and 
$\dot{M}=3.25\times 10^{-8}M_{\odot}yr^{-1}$ \citep{1998ApJ...492...323}. This is
our fiducial system. Two kinds of models are considered,
the first one is a $T_{0,fix}$ wall with constant surface temperature and the 
other is a wall, with a shape given by how the sublimation temperature depends
on density (see Figure~\ref{fig-shape}).

As noted in \S~\ref{sec-intro}, the modeling of interferometric 
observations primarily depends on two basic parameters: a typical distance to 
the emission region, and a typical temperature. In order
to compare these parameters with other works,
we present in Table~\ref{table-rad-temp} for all the models, either the 
$T_{0,fix}$ or the $T_{0,\rho}$ walls; the values of the minimum radius of the 
wall and the temperature at this location. For comparison, in 
Table~\ref{table-rad-temp} the parameters for the models of the 4 T Tauri stars
presented in \citet{2009ApJ...692...309} are shown. We are not looking a 
concordance between the values. As a matter of fact, we expect quite the 
opposite
because this set of values comes from two different approaches. Our modeling
produces these values from a detailed description of the wall formation. From
the other hand, \citet{2009ApJ...692...309} interpret their interferometric
observations with $R_{wall}$ and $T_{wall }$ as two free parameters of a 
simplistic model. In \citet{2009ApJ...692...309}'model, these parameters are
fitted without a model of the star or dust composition.

An important fact to notice here is that 
to obtain the interferometric image of the infrared observational data 
is required a prior knowledge of the observed structure. Then
the physical parameters extracted from infrared interferometric observations 
are calculated using a particular model; in other words, they are model 
dependent. These difficulties are not present in the
radio interferometry, where the large number of points in the visibility plane,
along with the high angular resolution and sensitivity obtained,
allows to apply the inverse Fourier transformation to get a real image of the 
object, that does not depend on a
particular model. In the case of NIR interferometry, the number of visibilities
is so low that we cannot invert the problem, thus, we depend on a model to 
extract the parameters of the system.

Summarizing, we should be careful to compare the set of parameters extracted in
this way with values given by a model independent of observations. 
\citet{2004ApJ...613...1049} uses 5 different models to interpret the 
interferometric observations at
$2.2\mu$m of 14 Herbig Ae/Be stars. These span geometries such as an envelope, 
disk or ring. Looking for the best fit allows to choose the model, however,
this do not completely guarantee that this is the right model. A qualitative 
comparison between the models presented here and the model consistent with 
observations can be done (see Table~\ref{table-rad-temp}).

\begin{table}[!t]\centering
  \setlength{\tabnotewidth}{0.5\columnwidth}
  \tablecols{3}
  \caption{Table of maximum $T_{wall}$ and minimum $R_{wall}$.}
  \label{table-rad-temp}
  \begin{tabular}{ccc}
    \toprule
    Model & $min(R_{wall})(AU)$ & $max(T_{wall})(K)$ \\ 
    \midrule
    $T_{0,fix}$($\dot{M}=1.625\times 10^{-8}M_{\odot}yr^{-1}$) & 0.0715 & 1400 \\
    $T_{0,fix}$($\dot{M}=3.25\times 10^{-8}M_{\odot}yr^{-1}$) & 0.079 & 1400 \\
    $T_{0,fix}$($\dot{M}=6.5\times 10^{-8}M_{\odot}yr^{-1}$) & 0.0915 & 1400 \\
    $T_{0,\rho}$($\dot{M}=1.625\times 10^{-8}M_{\odot}yr^{-1}$) & 0.0802 & 1402 \\
    $T_{0,\rho}$($\dot{M}=3.25\times 10^{-8}M_{\odot}yr^{-1}$) & 0.0801 & 1453 \\
    $T_{0,\rho}$($\dot{M}=6.5\times 10^{-8}M_{\odot}yr^{-1}$) & 0.0971 & 1430 \\
    Eisner et al. (2009) &  &  \\
    RY Tau & 0.16 & 1750 \\
    DG Tau & 0.18 & 1260 \\
    RW Aur & 0.14 & 1330 \\
    AS 205A & 0.14 & 1850  \\
    \bottomrule
  \end{tabular}  
\end{table}

For a $T_{0,\rho}$ sublimation wall model, we
note that the sublimation temperature ($T_{sub}$) depends on density 
\citep{1994ApJ...421...615}. Due to the fact that the density decreases with
height and that $T_{sub}$ increases with density, the modeled shape is convex. 
The lower denser parts of the wall are closer to the star; at high altitudes 
the density decreases, thus $T_{sub}$ decreases and the wall moves further out.
The density structure is given by a disk modeled 
with a detailed 2D numerical solution of the radiation transfer equations 
\citep{1998ApJ...500...411}. Models built
with this assumption are previously given by \citet{2005A&A...438...899} for
Herbig Ae stars.

\subsection{$T_{0,fix}$ WALL}
\label{sec-vertical}

A $T_{0,fix}$ wall is defined with a constant sublimation temperature in its
surface; here we take $T_{sub}=1400K$. For the luminosity of the typical low 
mass star, the
sublimation radius (equal to the wall location) is $R_{sub}=0.079\,AU$. 
The height of the wall is
taken as 5 times the pressure scale height \citep{2001ApJ...560...957}.
Note that the parameters required to define a $T_{0,fix}$ wall is $T_{sub}$ (or
$R_{sub}$) and a height (see Figure~\ref{fig-shape}).

\begin{figure}[!t]
  \includegraphics[width=\columnwidth]{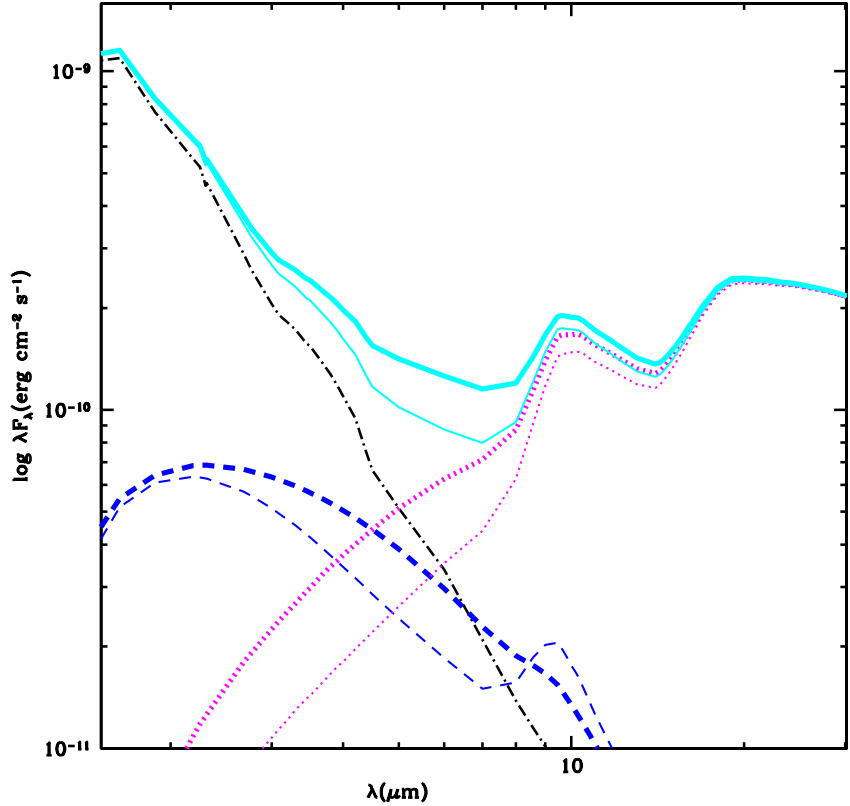}
  \caption{SEDs for the $T_{0,fix}$ model (thick line), and for 
the $T_{0,\rho}$ wall shaped by the dependence of the sublimation temperature on 
density (thin line) for the fiducial model. The dashed lines represent the wall
spectrum, the pointed lines the disk SED and the solid lines the total 
emission. The disk inclination is $\cos i=0.5$. }
\label{fig-SED-models}
\end{figure}

The SED of this model is presented in Figure~\ref{fig-SED-models}. In order to
do a fair comparison, we add to the wall and star spectra, a model of a disk 
with an inner radius equal
to the location of the wall and an outer radius equal to $100AU$. The disk
model is done using \citet{1998ApJ...500...411}.
The values of $T_{sub}$ and $R_{sub}$ for this model
are presented in Table~\ref{table-rad-temp}. Note that the wall spectrum does
not show the $10\mu$m silicate spectrum. Models changing the disk inclination
and the mass accretion rate are presented in \S~\ref{sec-curved}.

\subsection{$T_{0,\rho}$ WALL}
\label{sec-curved}

The shape of the $T_{0,\rho}$ wall is given by the fact that $T_{sub}$ depends on
 density \citep{1994ApJ...421...615}. The densest parts close to the midplane
have a $T_{sub}$ larger than the $T_{sub}$ in the upper layers of the disk, thus,
the former is closer to the star than the latter. The density vertical profile 
is taken from 2D axisymmetric disk models \citep{1998ApJ...500...411}.

\begin{figure}[!t]
  \includegraphics[width=\columnwidth]{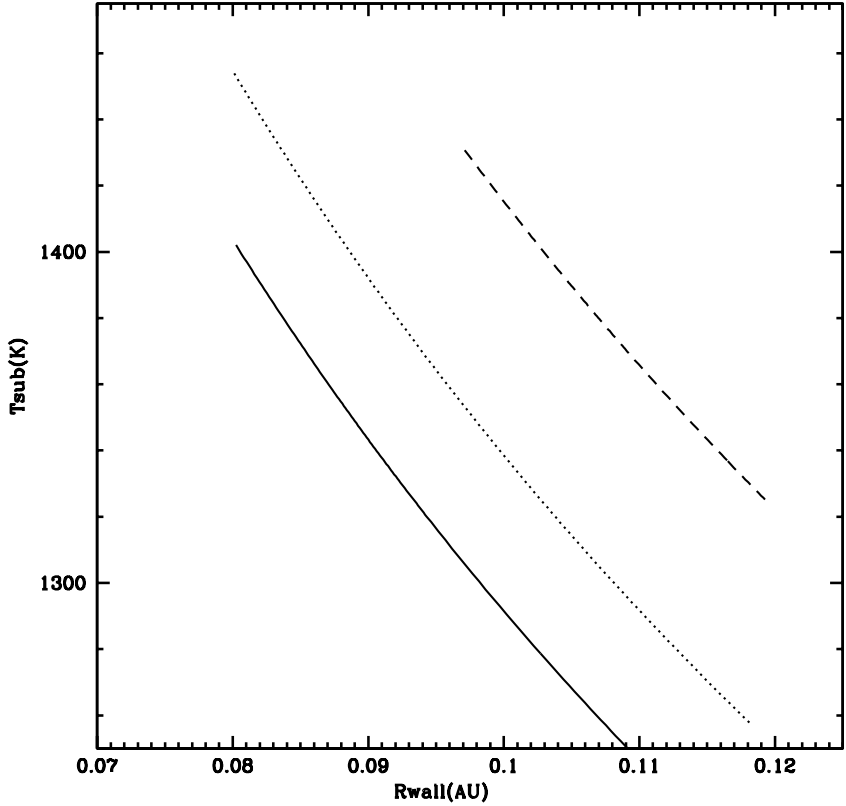}
  \caption{The sublimation temperature $T_{sub}$ along the $T_{0,\rho}$ wall 
surface 
is shown with a solid line for $\dot{M}=1.625\times 10^{-8}M_{\odot}yr^{-1}$,
with a pointed line for $\dot{M}=3.25\times 10^{-8}M_{\odot}yr^{-1}$, and with 
a dashed line for $\dot{M}=6.5\times 10^{-8}M_{\odot}yr^{-1}$. }
\label{fig-tsub}
\end{figure}

Figure~\ref{fig-tsub} presents a plot of $T_{sub}$ vs $R_{wall}$ for this case,
which shows a decreasing temperature with the radius, as one expects 
from a disk with a decreasing density with an increasing radius.  The shape of 
the wall is given in Figure~\ref{fig-shape}, either for $T_{0,\rho}$ or 
$T_{0,fix}$ walls for three values of $\dot{M}$. The value
$\dot{M}=3.25\times 10^{-8}M_{\odot}yr^{-1}$ corresponds to the fiducial model;
for comparison two models with half and twice this value are presented. Note
that the $T_{0,fix}$ wall location is given at the position where the temperature
is equal to $1400K$. The $T_{0,\rho}$ walls are located consistently outwards of
the $T_{0,fix}$ walls.

Figure~\ref{fig-SED-models} also presents the spectrum of the fiducial 
$T_{0,\rho}$
wall model. The model includes the outer disk, starting in this case at the
outer radius of the $T_{0,\rho}$ wall. The $T_{0,fix}$ wall emission is 
higher than the $T_{0,\rho}$ for $\lambda< 8\mu$m.
Also, the emission of the disk associated with the
$T_{0,fix}$ wall is higher because in this case the outer disk start at a smaller
radius. Putting these facts together, the SED of a system with a $T_{0,fix}$ wall
is noticeably higher than the model with a $T_{0,\rho}$ wall. This occurs mainly
between $3$ and $8\mu$m, resulting in differences between the IRAC colors, as
it is described in \S~\ref{sec-color}.

\begin{figure}[!t]
  \includegraphics[width=\columnwidth]{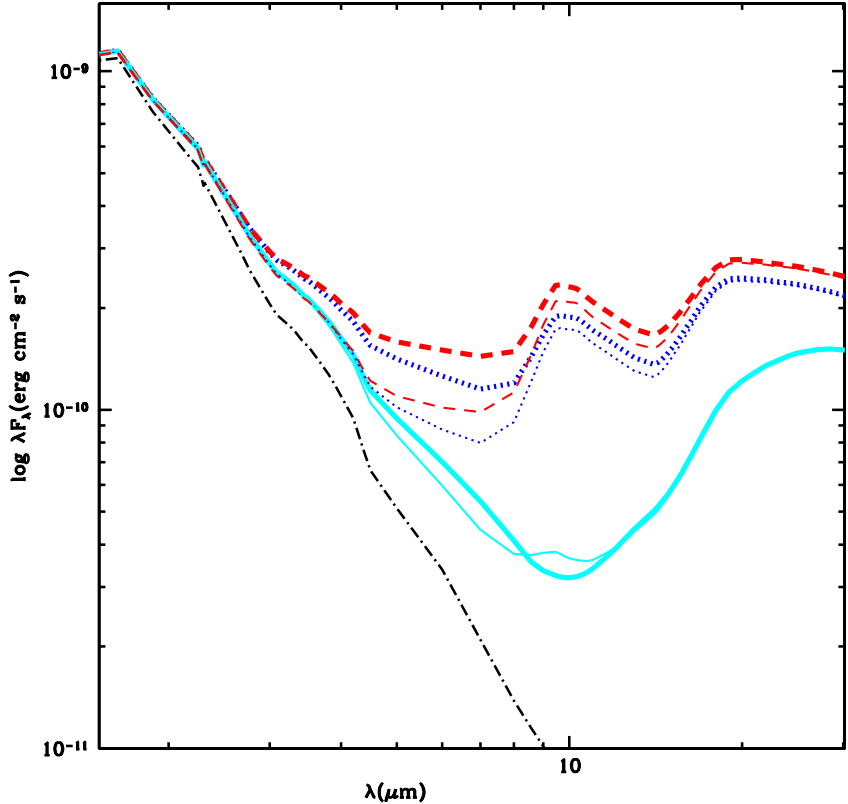}
  \caption{The spectrum of a $T_{0,fix}$ model (thick lines), and the 
$T_{0,\rho}$ model (thin lines). The inclination taken corresponds to 
$\cos i=0.3$ (solid lines), $\cos i=0.5$ (pointed lines), and $\cos i=0.7$ 
(dashed lines).} 
\label{fig-SED-incl}
\end{figure}

Neither the mass accretion rate or the inclination are parameters not well 
defined or even not defined at all for real systems. In order to see models
with different values for these parameters and to be sure that the models do
not overlap, we present another set of models.
Figure~\ref{fig-SED-incl} presents for the fiducial model, the behavior of the
spectrum with inclination, either for $T_{0,fix}$ or $T_{0,\rho}$ models. It is 
important to note that for each type of wall, the emission decreases as the
inclination increases. Note that for a higher inclination, the surface of the 
disk in the sky plane decreases, which naively means a lower emission.
Besides, note that at $\lambda<8\mu$m, a $T_{0,fix}$ wall model with $\cos i=0.5$
($60^{\circ}$) emits more than a $T_{0,\rho}$ wall with $\cos i=0.7$ 
($45^{\circ}$).
It is important to point out that the shape of the SED for these two models is
different, thus, in principle a change in inclination is not able to match 
models with a
$T_{0,\rho}$ and a $T_{0,fix}$ wall, thus, in this way a modeler should be able 
to distinguish between these two scenarios, of course depending on the 
resolution and precision of the spectrum.

\begin{figure}[!t]
  \includegraphics[width=\columnwidth]{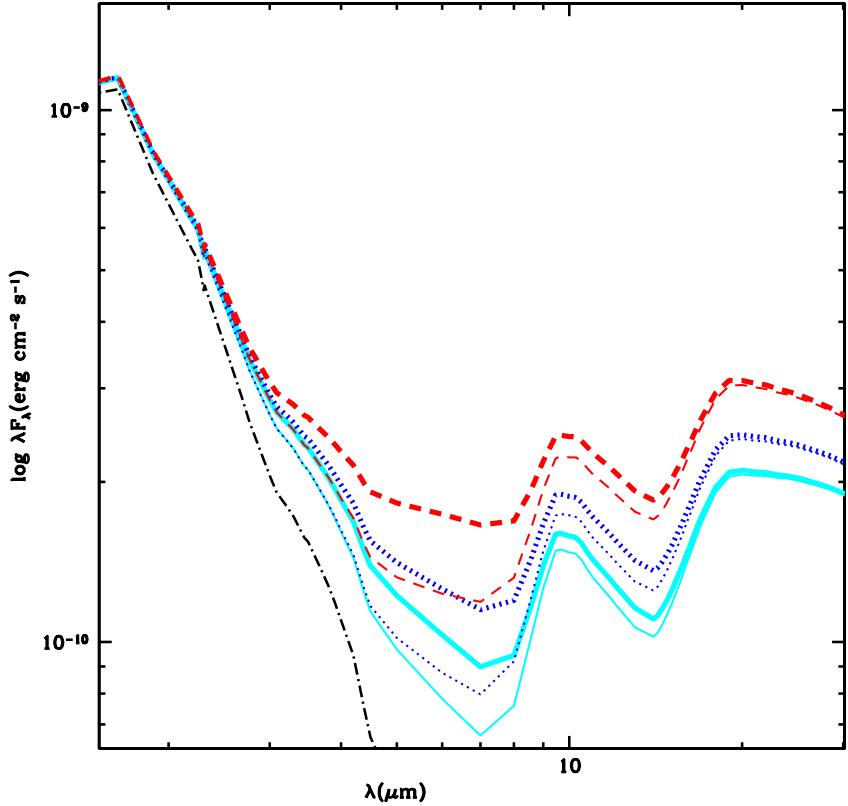}
  \caption{The spectrum of a $T_{0,fix}$ model (thick lines), and the 
$T_{0,\rho}$ model
(thin lines) in terms of mass accretion rate. The mass accretion rate taken 
corresponds to $\dot{M}=1.625\times 10^{-8}M_{\odot}yr^{-1}$ (solid line),
$\dot{M}=3.25\times 10^{-8}M_{\odot}yr^{-1}$ (pointed line), and  
$\dot{M}=6.5\times 10^{-8}M_{\odot}yr^{-1}$ (dashed line).}
\label{fig-SED-mdot}
\end{figure}

In Figure~\ref{fig-SED-mdot}, plots for the SED in terms of $\dot{M}$ are 
presented. It is expected that either for the $T_{0,fix}$ or $T_{0,\rho}$
wall models, increasing $\dot{M}$ means a larger flux, and this do not mean 
just to move the SED by a constant amount, but a change in the shape of the
curve.
Note that the emission for a $T_{0,fix}$ wall model with
$\dot{M}=3.25\times 10^{-8}M_{\odot}yr^{-1}$ is larger than the emission from a
$T_{0,\rho}$ wall model with $\dot{M}=6.5\times 10^{-8}M_{\odot}yr^{-1}$ for
$\lambda<7\mu$m. This is an example of the differences between $T_{0,\rho}$ and
$T_{0,fix}$ wall models, which correspond to changes in the IRAC colors as can be
seen in \S~\ref{sec-color}.

\section{COLOR INDICES}
\label{sec-color}

In \citet{2004ApJS...154...363} there is a comparison of IRAC colors between 
models with $T_{0,fix}$ walls and a sample of young stellar objects. Here, we
note a difference in the SED between the models with $T_{0,fix}$ and $T_{0,\rho}$ 
walls in the wavelength range defined by $3$ and $8\mu$m. This means that the
IRAC colors will show variations between the models. 
The emission of the $T_{0,fix}$ wall that they use is
a blackbody at the typical sublimation temperature of $1400$K. Note that the
emission from our model for the $T_{0,fix}$ wall comes from the atmosphere, with
the surface at $1400$K, but the inner parts a little bit cooler. This means that
ours and theirs colors for models with the same parameters should not be the
same. The [3.6]-[4.5] color range coincides between \citet{2004ApJS...154...363}
and this work. The [5.8]-[8.0] color value associated to our models is 
consistently larger than the values in \citet{2004ApJS...154...363}. Note that
even for the $T_{0,fix}$ walls, the emission comes from material in a range of
temperatures, in particular at temperatures lower than $1400$K, which is the 
temperature taken in \citet{2004ApJS...154...363}. This should change in 
particular the [5.8]-[8.0] color. The zero point magnitudes are taken as in
\citet{2005ApJ...629...881}, consistent with a Vega-based IRAC magnitude 
system.

Recently, \citet{2010ApJS...188...75} estimates spectral indices in the range
between $6$ and $31\mu$m for disk models, using $T_{0,fix}$ walls at $1400$K.
As noted here, a change in the emission due to a $T_{0,\rho}$ wall instead of a
$T_{0,fix}$ wall will modify the spectral indices, however, due to the dispersion
of the values in \citet{2010ApJS...188...75}, including a $T_{0,\rho}$ wall will
not change their main conclusions. On the other hand, when one tries to fit
specific objects, taking into account a $T_{0,\rho}$ wall should be an 
unavoidable modeling requirement.

\begin{figure}[!t]
  \includegraphics[width=\columnwidth]{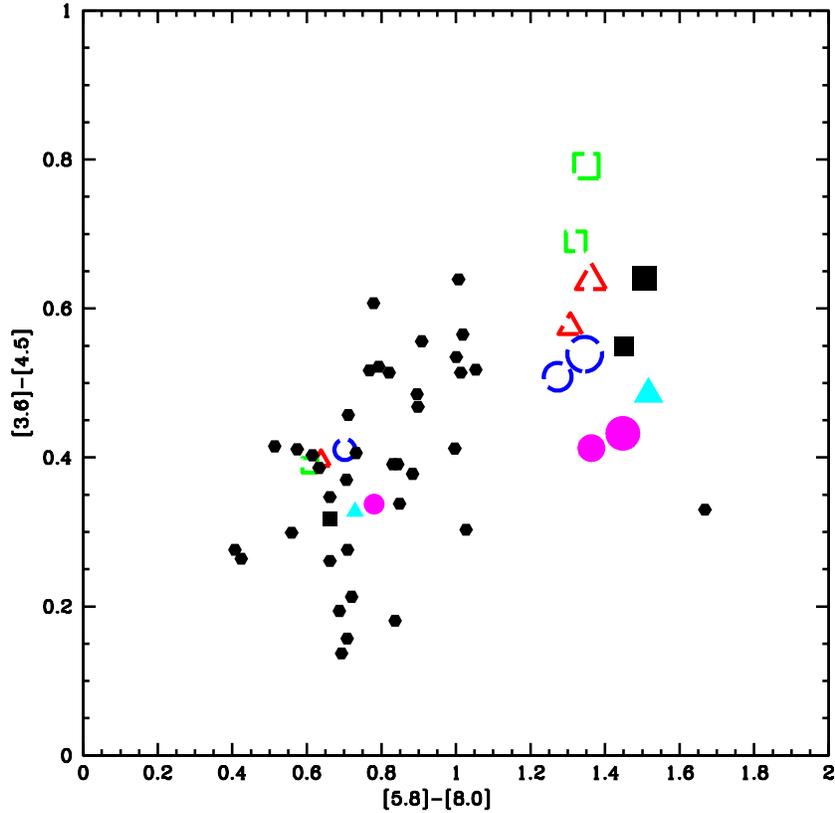}
  \caption{IRAC colors for models including star, wall and disk. The models 
with a $T_{0,fix}$ wall or a $T_{0,\rho}$ wall are presented with open and filled 
symbols, respectively. Models for the following mass accretion rates are shown: 
$\dot{M}=1.625\times 10^{-8}M_{\odot}yr^{-1}$ (circles),
$\dot{M}=3.25\times 10^{-8}M_{\odot}yr^{-1}$ (triangles), and
$\dot{M}=6.5\times 10^{-8}M_{\odot}yr^{-1}$ (squares). The smaller size symbols
correspond to an inclination given by $\cos i=0.3$, the medium size 
correspond to $\cos i=0.5$, and the larger size to $\cos i=0.7$. The black 
points correspond to the sample of Class II pre-main-sequence objects in
\citet{2005ApJ...629...881}.}
  \label{fig-color}
\end{figure}

In Figure~\ref{fig-color}, the IRAC color-color diagram is presented for 
$T_{0,fix}$ and $T_{0,\rho}$
wall models. Models with $\dot{M}=(1.625,3.25,6.5)\times 10^{-8}M_{\odot}yr^{-1}$,
and $\cos i=(0.3,0.5,0.7)$ are presented. The $\cos i=0.3$ $T_{0,fix}$ wall 
models
are clearly located around $[3.6]-[4.5]\sim 0.4$ and $[5.8]-[8.0]\sim 0.65$. 
The less inclined models moves up and right in this plot, either for the 
$T_{0,fix}$ or $T_{0,\rho}$ walls models. The $T_{0,\rho}$ wall models compared 
with respect
to their $T_{0,fix}$ counterparts, decrease the $[3.6]-[4.5]$ color and increase
the $[5.8]-[8.0]$ color. The trend of the colors in the plot changing $\dot{M}$
and $\cos i$, allow us to confidently say that there is no overlapping of the
$T_{0,\rho}$ and $T_{0,fix}$ wall models. Thus, one cannot confuse a $T_{0,fix}$ 
and a $T_{0,\rho}$ wall model when changing these two parameters.

In the last paragraph, we stress that the various models are located in
specific regions of the color-color diagram. Specifically, looking at the 
models in Figure~\ref{fig-color}, there is no degeneracy with the inclination.
At first sight, this looks as a surprising result, because it is usually 
assumed that the inclination can work as a tuning parameter for the emitted
flux. Notwithstanding, it is highly probable that we can tune one of the IRAC
magnitudes, but it is difficult to think that one can tune the four at the same
time. In favor of this, this is not just a geometrical problem, because the 
flux comes from a temperature distribution that depends on the position in the
wall atmosphere, different between the models.

In order to compare the modeled color indices with real systems, in
Figure~\ref{fig-color} the black points correspond to the Class II objects of 
the \citet{2005ApJ...629...881}'s sample of Taurus. A direct comparison point
out that the objects are consistent with high inclination disks. However, a
definitive answer requires a modeling of the systems one by one, which is
beyond the scope of this paper. 

Summarizing, the results presented in Figure~\ref{fig-color} lead us to 
the conclusion that the models are not degenerate for the parameters used. In
particular, if one can observationally fit the inclination (i.e. using
interferometric images), a comparison of the observed and modeled colors will
allow to choose between the models.

\section{SUMMARY AND CONCLUSIONS}
\label{sec-summary}

In this paper, for simplicity the heating from inner layers is not taken into 
account. This is responsible to move
outwards the rim \citep{2009A&A...506...1199}, and also to change the shape of 
the wall. Here, we describe the wall as a stationary well
defined surface, in spite of the results of \citet{2009A&A...506...1199}. They  
developed a code that find the sublimation wall with a detailed description of
the sublimation process: the wall (dust-gas boundary) is a non-stationary
diffuse region. Our reason for this assumption is that the calculation of the
emission for a time dependent diffuse region is a very complex issue, which 
depends on a lot of unknowns. For the lack of this knowledge, to pursue further
in the improvement of the model is not worthed at this moment.  
Our goal is to detect the differences on the SED, in particular for 
changes on the vertical geometry assumption, which we think are not 
conspicuously modified due to these assumptions.

There is a difference between the $T_{0,fix}$ and $T_{0,\rho}$ 
walls SEDs in the wavelength range between $\lambda =3$ and $8\mu$m. Due to 
this, the near-infrared colors calculated with a disk plus a $T_{0,fix}$ wall
(commonly used) and a disk plus a $T_{0,\rho}$ wall differ. However, when 
analysing
sets of spectra \citep{2004ApJS...154...363,2005ApJ...629...881} using the
IRAC color-color diagram, this 
difference do not change the conclusions previously presented, due to the
dispersion of models and observations in this plot. For the modeling of real 
objects spectra, the decision of which wall model to take, should be done 
carefully. 

The main conclusions are summarized next.

\begin{itemize}

\item A $T_{0,fix}$ wall is closer than a $T_{0,\rho}$ wall. $R_{wall}$ changes 
between $T_{0,fix}$ or $T_{0,\rho}$ wall models, at most by $10\%$.

\item The emission of a $T_{0,fix}$ wall is larger than the emission of a 
$T_{0,\rho}$ wall.

\item For each type of wall, the emission decreases as the inclination 
increases.

\item A change in inclination is not able to match models with a $T_{0,\rho}$ 
and a $T_{0,fix}$ wall.

\item The $T_{0,fix}$ wall do not show the $10\mu$m silicates band (see 
figure~\ref{fig-SED-models}).

\item The disk for models with both types of walls shows the silicates band.
For this reason, in the spectrum of the star-wall-disk system is always present
this feature.

\item In IRAC color-color diagram, the $T_{0,fix}$ wall
models are located in a region different with respect to the $T_{0,\rho}$ wall 
models. Between the $T_{0,\rho}$ and $T_{0,fix}$ wall models, the [3.6]-[4.5] and
[5.8]-[8.0] colors changes in around $20\%$ and $10\%$, respectively.
\end{itemize}

\end{document}